# Multiphase Unified Flash Analysis and Stability Algorithm
## (MUFASA V1.0):
## An Open-Source Software

## "The Manual"

**Ali Zidane**

Independent Researcher*

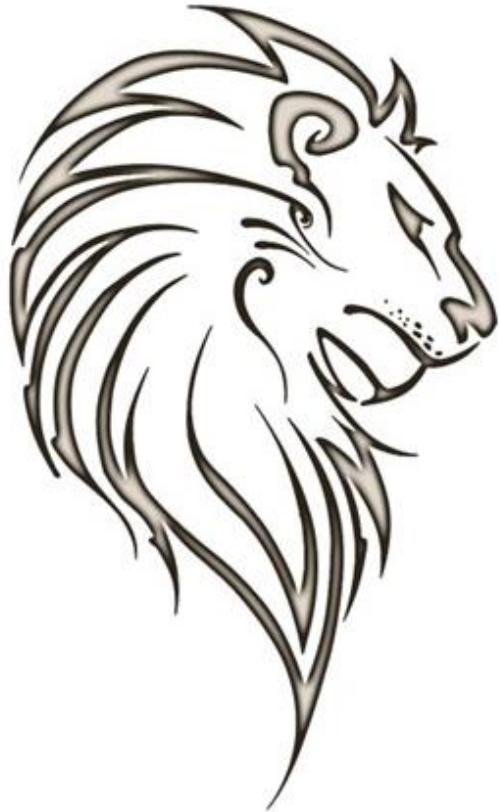

*Now working at the national laboratory, USA

# 1. Introduction

The software in this package performs two-phase, three-phase and four-phase split calculations. The calculations are based on the Peng-Robinson equation of state (PR-EOS). The four-phase calculations are based on a geometrical solution in 3D for Rachford-Rice equations.

In order to perform multiphase split calculations, a set of initial guesses in both stability testing and phase-split calculations are used. The initial guesses are used to locate the stationary points of tangent plane distance (TPD). The TPD is widely used in stability analysis. A given phase is stable if TPD is positive, otherwise it is unstable (Boyan et al. 2013).

The software is developed in Python with a user-friendly Graphical User Interface. The following provides a description of the interface and the steps needed to run a simulation. As will be shown the user can readily set and run a multiphase split simulation for thousands of different pressure-temperature conditions. The software is available to download at the (https://github.com/alizidane01/MUFASA_Public).

# 2. A 3D geometrical approach

The simulator uses a geometric approach to perform multiphase (up to four phase) split calculations. The geometric approach solves for Rachford-Rice (RR) equations given below (Imai et al. 2019, Mohebbinia et al. 2013):

$$\begin{cases} RR_y(\beta_y, \beta_z, \beta_t) = \sum_i \frac{(K_{yi}-1)n_i}{1 + \beta_y(K_{yi} - 1) + \beta_z(K_{zi} - 1) + \beta_t(K_{ti} - 1)} = 0 \\ RR_z(\beta_y, \beta_z, \beta_t) = \sum_i \frac{(K_{zi}-1)n_i}{1 + \beta_y(K_{yi} - 1) + \beta_z(K_{zi} - 1) + \beta_t(K_{ti} - 1)} = 0 \\ RR_t(\beta_y, \beta_z, \beta_t) = \sum_i \frac{(K_{ti}-1)n_i}{1 + \beta_y(K_{yi} - 1) + \beta_z(K_{zi} - 1) + \beta_t(K_{ti} - 1)} = 0 \end{cases} \quad (1)$$

where $\beta$ is the mole fraction of a phase ($y$, $z$, $t$) and $K_i$ is the equilibrium ratio of component $i$ with respect to a reference phase and $n_i$ is the overall mole fraction of component $i$.

If a direct approach (e.g. NR method) is used to solve the above equations, a good initial guess is needed for convergence. A geometric approach, however, overcomes this limitation.

When the mixture is in four-phase, three unknowns (mole fractions of the phases) need to be calculated. The mass conservation equation gives the fourth unknown. A geometric solution of the RR equations with a provided equilibrium ratios is based on the intersection of the three equations in (1) in 3D. The constraints of mass conservation to unity in 2 and 3 phases, enforce the solution to be in a tetrahedron. Fig.1 shows the iso-surfaces of the three equations in (1) that satisfy the conditions $RR_y = RR_z = RR_t = 0$. The conditions and composition of the mixture that are used for this figure will be given in the example in the following sections. Fig.2 shows their intersection. The intersection of the three satisfied constraints provides the solutions for the mole fractions of the phases. The $K$-values that appear in (1) that are used in stability analysis and phase split for the different state mixtures are given in Table 1.

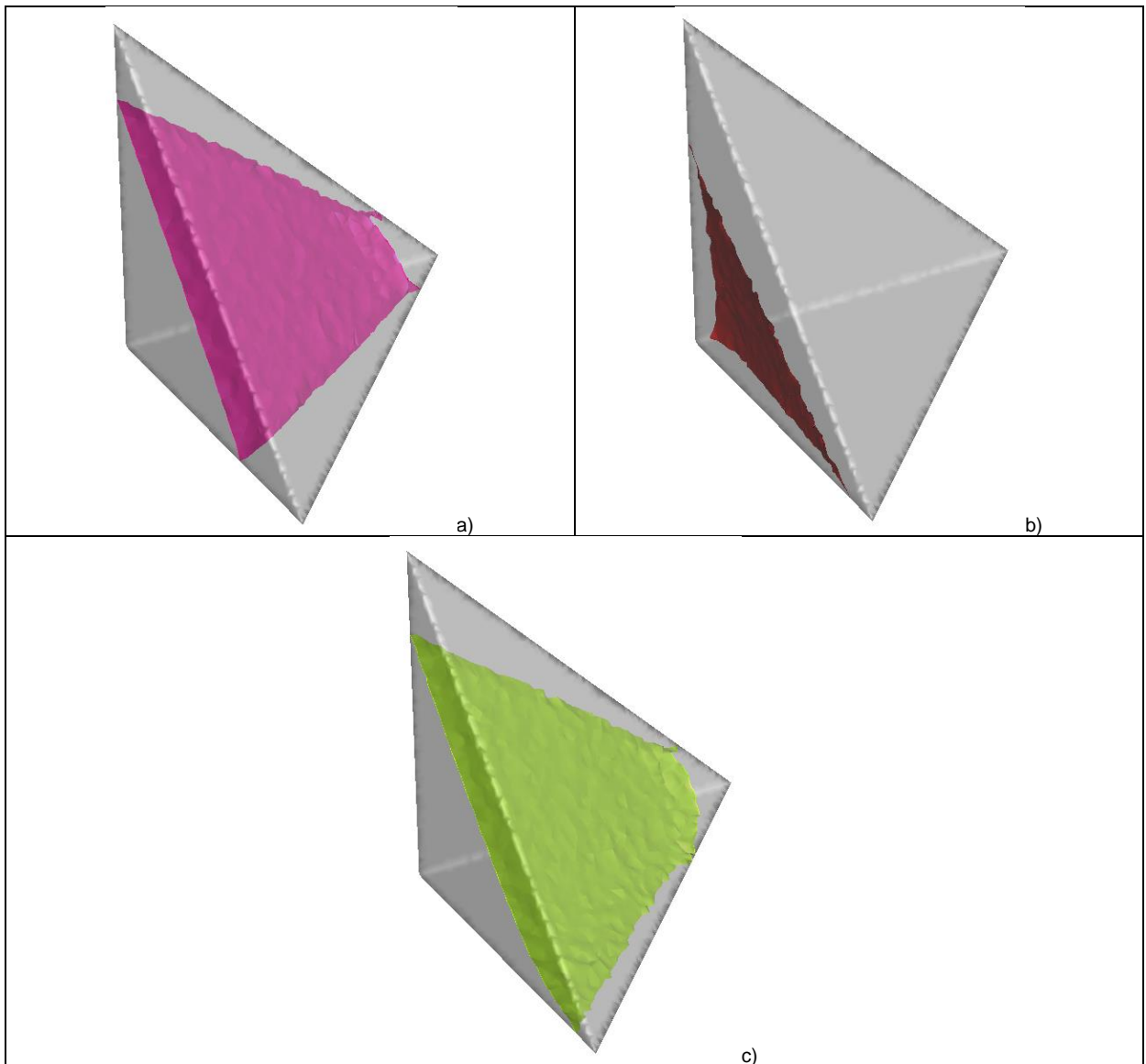

**Fig.1: Iso-surfaces that satisfy the conditions in equation (1) for RRy (a), RRz (b), and RRt (c)**

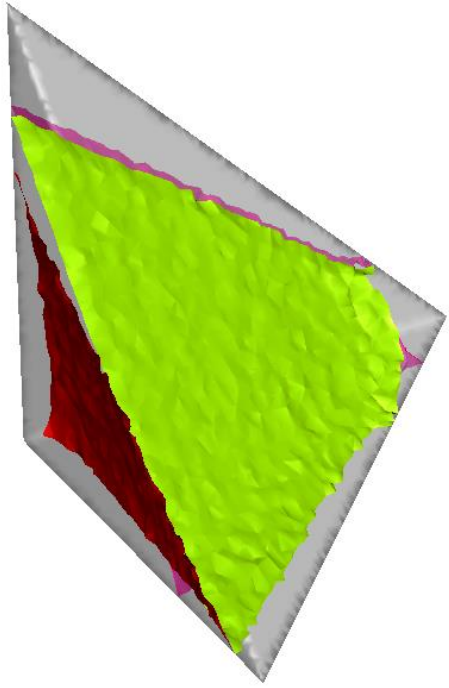

**Fig.2: Intersection of the three iso-surfaces (i.e. solutions) of the three conditions in equation (1)**

| State of the mixture | K-values |
|---|---|
| One-phase stability (1-stab) | $K_i^*, K_i^w, 1/K_i^w, \sqrt[3]{K_i^w}, 1/\sqrt[3]{K_i^w}$ |
| Two-phase flash (2-F) | $K_i^{1-stab}, 1/K_i^{1-stab}, K_i^w, 1/K_i^w$ |
| Two-phase stability (2-stab) | $K_i^*, K_i^w, 1/K_i^w, \sqrt[3]{K_i^w}, 1/\sqrt[3]{K_i^w}, K_i^{1-stab}, 1/K_i^{1-stab}, K_i^{2-F}, 1/K_i^{2-F}$ |
| Three-phase flash (3-F) | $K_i^w, 1/K_i^w, K_i^{1-stab}, 1/K_i^{1-stab}, K_i^{2-F}, 1/K_i^{2-F}, K_i^{2-stab}, 1/K_i^{2-stab}$ |
| Three-phase stability (3-stab) | $K_i^*, K_i^w, 1/K_i^w, \sqrt[3]{K_i^w}, 1/\sqrt[3]{K_i^w}, K_i^{1-stab}, 1/K_i^{1-stab}, K_i^{2-F}, 1/K_i^{2-F}, K_i^{3-F}, 1/K_i^{3-F}$ |
| Four-phase flash (4-F) | $K_i^w, 1/K_i^w, K_i^{1-stab}, 1/K_i^{1-stab}, K_i^{2-F}, 1/K_i^{2-F}, K_i^{2-stab}, 1/K_i^{2-stab}, K_i^{3-F}, 1/K_i^{3-F}$ |

$$K_i^* = \begin{cases} \frac{0.99}{x_{i,r}} \; ; \; i = 1,..n_c \\ \frac{0.01}{\frac{n_c}{x_{j,r}}} \; (j \neq i) \end{cases} \quad ; \quad K_i^w = \frac{p_{c,i}}{p} \exp\left[5.37(1 + w_i)\left(1 - \frac{T_{c,i}}{T}\right)\right]$$

$p_{c,i}, T_{c,i}, w_i: critical\ pressure, crtical\ tempeature, acentric\ factor$

**Table1: K-values used in the different states of the mixture**

## 3. The Graphical user interface

I provide in this section a description of the graphical interface. I will be using as an example the four-phase mixture reported in Paterson et al. (2017) as a reference solution for validation. The mixture contains $CO_2/C_1/C_2/C_3/H_2S$ and phase split is performed over range of pressure from 1 to 10 bar and a range of temperature from 130 to 150 K.

Following the steps in this manual the user should be able to reproduce the multiphase contour plot reported in (Paterson et al. 2017) based on the calculation of thousands of pressure and temperature conditions.

Double clicking the file "Mufasa.exe" opens the main window of the software as shown in Fig.3.

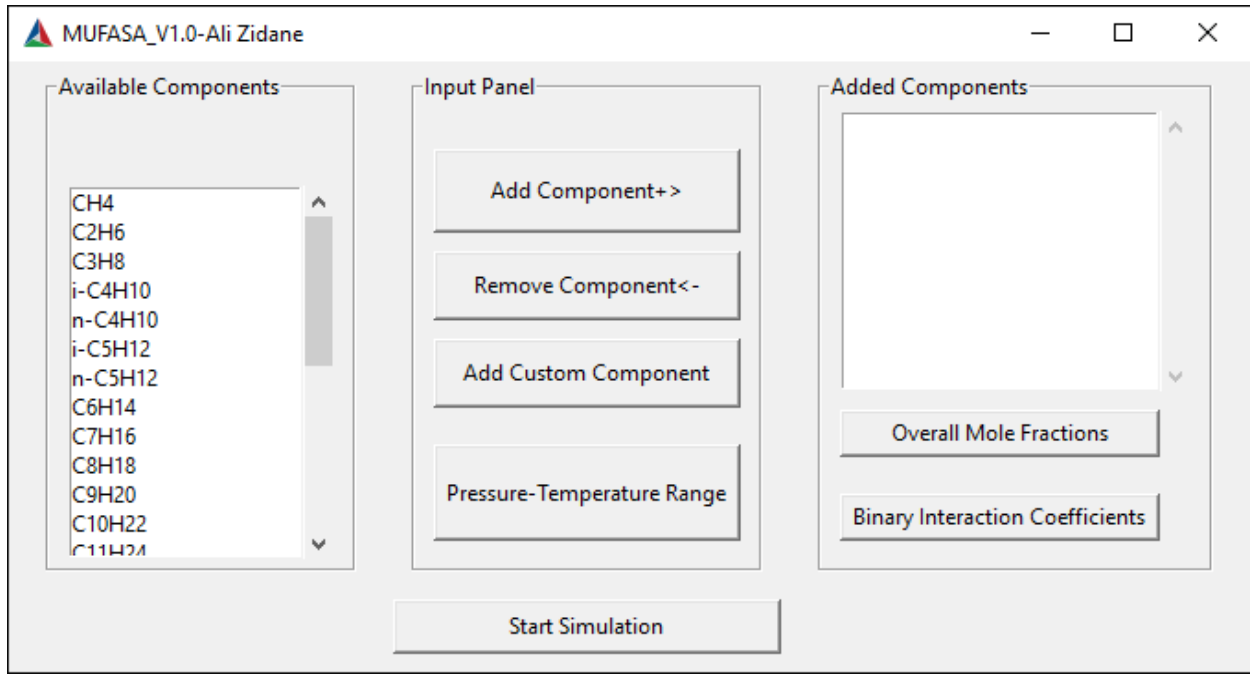

Fig.3: Main window of the software

The above figure shows three main panels: left, middle and right. The left panel contains the available components the user can choose from. In the release package, an accompanying file named "Data.xlsx" contains the components and their properties that are shown to be available in the left panel. One could update this file by adding more components to the file following the same format of the file or click the button "Add Custom Component" in the middle panel of the software. Note that molar mass is in [g/mole], Tc in [K], Pc in [bar] and Vc [$m^3$/kg].

- To add a component to the mixture, the user needs to select the component from the left panel and click the "Add Component" button. The component should now appear in the right panel. Fig.4 shows the selection of a mixture of $CO_2/C_1/C_2/C_3/H_2S$.
- If a component is added by mistake, the user can select the component from the "Added Components" window and click "Remove Component" to remove it from the mixture.

- Once the components have been selected, the user needs to click the "Overall Mole Fractions" to enter the mole fraction for each component in the mixture. Fig.5 shows the mole fractions for the components of the mixture discussed above. Once the values are added click "Add Fractions".

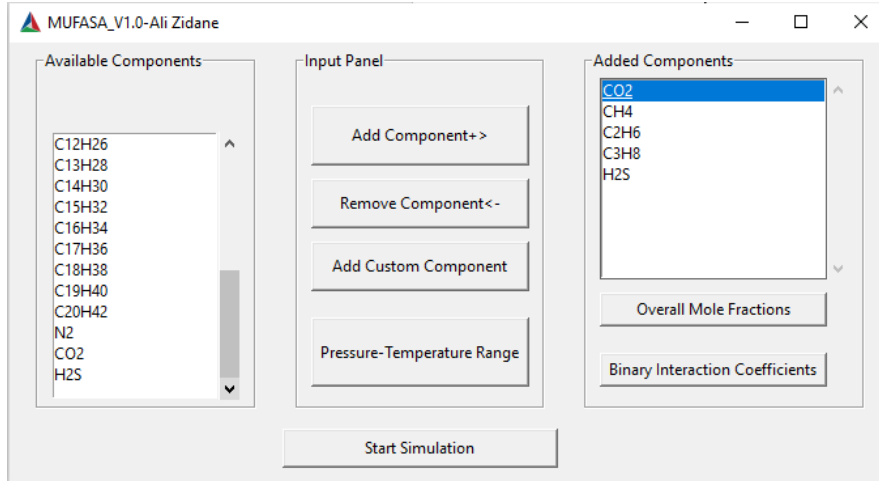

Fig.4: Selection of the mixture

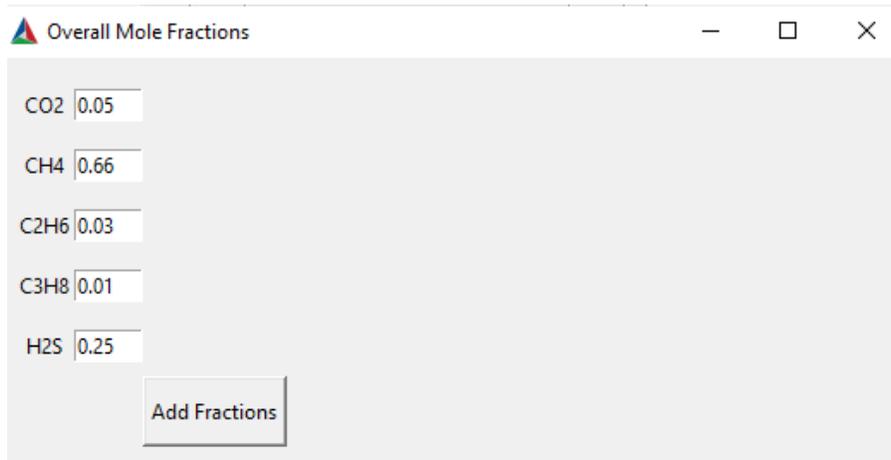

Fig.5: Overall mole fractions

- After the mole fractions have been entered, the user needs to click "Binary Interaction Coefficients" (BIC). Since the BIC matrix is symmetric, only the lower half will be shown. Default values are zeros. For this example, we are going to change the BIC matrix values as shown in Fig.6.

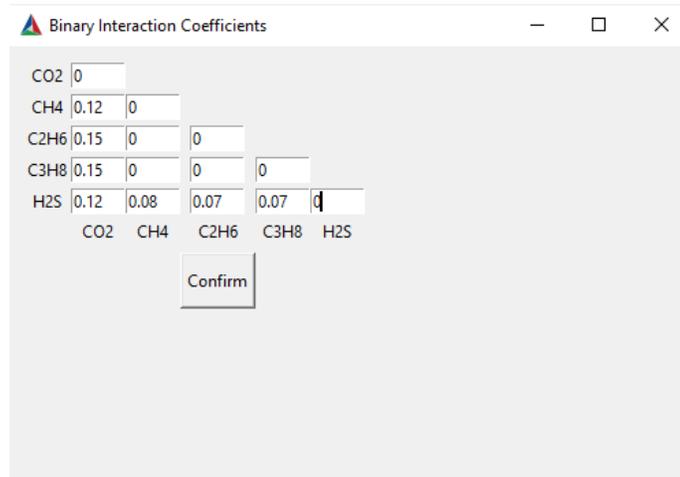

Fig.6: Binary Interaction Coefficients lower half matrix

- The last input that is needed to run a simulation is the range of pressure and temperature the user wants the calculations to be performed on. Clicking the "Pressure-Temperature Range" button in the middle panel. Fig.7 shows the PT range that will be used in this simulation example (1-10 bar for Pressure and 130-150 K for Temperature). In addition to the PT range, the user needs to enter the number of calculation points for each of two parameters. The software will therefore split the parameter range over the number of points equally. In this example a 100 calculation points for pressure and a similar number for temperature will be used. A total of 10-thousand calculations will be performed. For the pressure the increment of calculation points in this example will be 0.1 bar and for the temperature it will be 0.2 K. Lastly, the "Enter P-T range" button should be clicked.

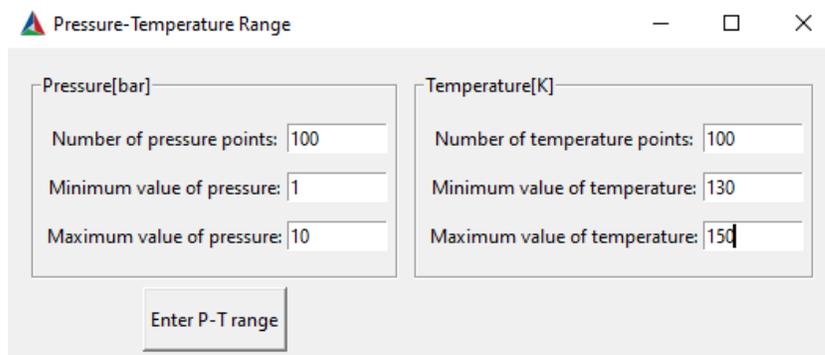

Fig.7: Pressure-Temperature range

Once all the above is completed, the user needs to click the "Start Simulation" button in the middle panel. After the simulation is completed, a message box will appear as shown in Fig.8. Click to close all windows.

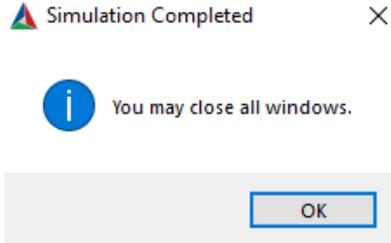

Fig.8: Message box declaring end of simulation

## 4. Demonstration

The software outputs two files: "results.dat" and "results_phases.csv". The first file (results.dat) includes the phase split calculations for all the input points and the mole fraction of components among the phases and other details. A sample of data from that file is shown below. This condition corresponds a pressure value of 3.09 bar and a temperature value of 131.21 K. The mixture at those conditions is in four-phase.

```
P (bar) =    3.090909
T (K)   =  131.212121
Mixture is in  "FOUR-PHASE"

FOUR-PHASE Flash computation
-----------------------------

idx     feed        Phase1        Phase2        Phase3        Phase4      comp
----    ----        ------        ------        ------        ------      ----
 1     0.0500      0.00262443    0.05638951    0.05727660    0.89955197    CO2
 2     0.6600      0.99617338    0.74054772    0.02677896    0.03706214    CH4
 3     0.0300      0.00085077    0.11356870    0.00490258    0.00188622    C2H6
 4     0.0100      0.00000627    0.03950353    0.00059362    0.00016144    C3H8
 5     0.2500      0.00034515    0.04999054    0.91044824    0.06133822    H2S

                                  Phase1        Phase2        Phase3        Phase4
                                  ------        ------        ------        ------
MOLAR FRACTION:                   0.46958       0.24908       0.25925       0.02209
VOLUME FRACTION:                  0.98874       0.00590       0.00493       0.00043
ZFACTOR:                          0.92994       0.01046       0.00841       0.00854
MW(g/mol):                       16.09190      21.19510      34.16867      42.32881
MASS DEN(Kg/L):                   0.00490       0.57393       1.15169       1.40371
MOLE DEN(mol/L):                  0.30466      27.07857      33.70601      33.16208
```

Fig.9: Sample of the calculation results

The second output file is "results_phases.csv". This file contains the number of phases for each of the calculation point. It is provided in a matrix form so the user can plot it in 2D for a better visualization. Fig.10 shows the contour plot for the distribution of the number of phases over the calculation range. Results are compared to those reported in (Paterson et al. 2017) and a good agreement is observed as shown in Fig.10.

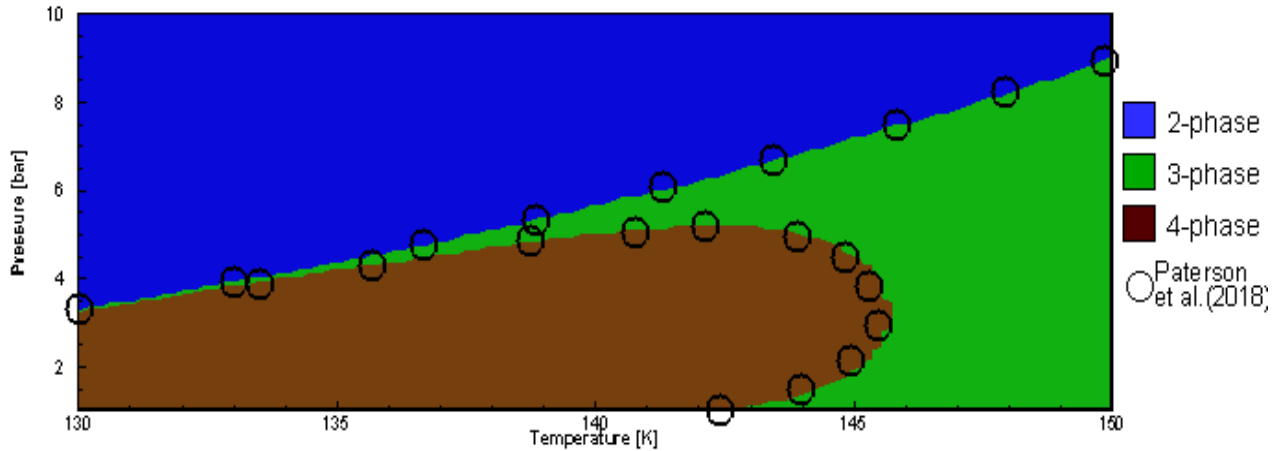

**Fig.10: 2D plot showing the distribution of the total number of phases on the PT calculation range**

Software available through the following link:
https://github.com/alizidane01/MUFASA_Public